\documentclass{JHEP}
\usepackage{graphicx,amsmath,amssymb}
\usepackage{epsfig,multicol}
\usepackage{epsf}
\usepackage{epstopdf}
\input{epsf.sty}

\newcommand{\ba}{\begin{eqnarray}}
\newcommand{\ea}{\end{eqnarray}}
\def\be{\begin{equation}}
\def\ee{\end{equation}}

\def\ap{\alpha^{\prime}}
\def\D{\bar D}

\title{Brane Inflation :   String Theory viewed from the Cosmos
\footnote{To appear in "String theory and fundamental interactions", Eds. M. Gasperini and J. Maharana (Lectures Notes in Physics, Springer-Verlag, Heidelberg, 2007), published in occasion of the 65th birthday of Gabriele Veneziano.}
}
\author{S.-H.~Henry Tye\footnote{Electronic mail:
tye@lepp.cornell.edu}
\\Newman Laboratory for Elementary Particle Physics, Cornell University, 
Ithaca, NY 14853, USA}

\abstract{Brane inflation is a specific realization of the inflationary universe scenario in the early universe within the brane world framework in string theory. The naturalness and robustness of this realistic scenario is explained. Its predictions on the cosmological observables in the cosmic microwave background radiation, especially possible distinct stringy features, such as large non-Gaussianity or large tensor mode that deviates from that predicted in the slow roll scenario, are discussed. Stringy KK modes as hidden dark matter is also a possibility. Another generic consequence of brane inflation is the production of cosmic strings towards the end of inflation. These cosmic strings are nothing but superstrings stretched to cosmological sizes. The properties of these cosmic superstrings and their subsequent cosmological evolution into a scaling network open up their possible detections in the near future, via cosmological, astronomical and/or gravitational wave measurements. At the moment, cosmological data is already imposing strong constraints on the details of the scenario. 
Finding distinctive stringy signatures in cosmological observations will go a long way in revealing the
specific brane inflationary scenario and validating string theory as well as the brane world picture. Precision measurements may even reveal the structures of the flux compactification. Irrespective of the final outcome, we see that string theory is confronting data and making predictions. \\
\\
\\

 ~~~~~~~~~~   In celebration of the 65{\it th} birthday of Gabriele Veneziano, teacher and friend.
}

\begin{document}

\maketitle

\section{Introduction}

It is believed by many that superstring theory is the fundamental theory of all matter and forces, 
including a consistent quantum gravity sector. In fact, it is the only known theory that incorporates general relativity in a quantum mechanically consistent way around the near Minkowski spacetime
 that describes our universe today. The theory is also extraordinarily intricate, revealing numerous 
deep and rich mathematical and physical structures.
However, the string scale is believed to be so high that it is almost hopeless to find stringy 
signatures at any high-energy experiments in the conceivable future. Since such a high energy 
scale was probably once reached in the early universe, it is natural to look for stringy signatures 
in cosmology. Looking towards the sky for information and tests on fundamental physics has a long tradition. This follows the route taken by, for example, the discovery of Newton's gravitational force law and Einstein's theory of general relativity.

The inflationary universe was proposed to solve a number of fine-tuning problems such as the flatness problem, the horizon problem and the defect problem \cite{Guth:1980zm}. Besides providing an origin for the hot big bang (the ultimate free lunch), its prediction of an almost scale-invariant density perturbation power spectrum (which is responsible for structure formation in our universe) has received strong observational support from the temperature fluctuation and polarization in the cosmic microwave background radiation (CMBR), e.g., COBE \cite{Smoot:1992td} and WMAP \cite{Spergel:2006hy}. However, the origin of the key ingredients of the inflationary scenario, namely, the scalar field known as the inflaton and its potential, remain undetermined. In this sense, the inflationary universe scenario is considered by many to be a paradigm or framework, not quite a theory. As the cosmological data keeps improving in a very impressive fashion, it becomes urgent to find a specific model that has a solid theoretical foundation.
 
If string theory is the theory of everything, we should be able to find a natural inflationary scenario there. This will allow us to identify the inflaton and its properties, while at the same time cosmological measurements will help us to determine the precise stringy description of our universe. With some luck, we may even find distinct stringy signatures in this framework in the cosmological data to confirm our faith in the theory. Since the inflationary scale turns out to be comparable to the string scale, such an investigation is clearly very worthwhile. If the scenario is natural, one should be able to explain why many e-folds of expansion is generic (without fine-tuning). A good test requires the scenario/model to be over-constrained, that is, the number of measurements should eventually exceed the number of parameters in the model. We shall explain how (and in what sense) brane inflation, a specific realization of the inflationary universe scenario in the early universe within the brane world framework in string theory, satisfies these 2 criteria; that is, it is both natural and testable.

\begin{figure}
\begin{center}
\includegraphics[width=0.8\textwidth,angle=0]{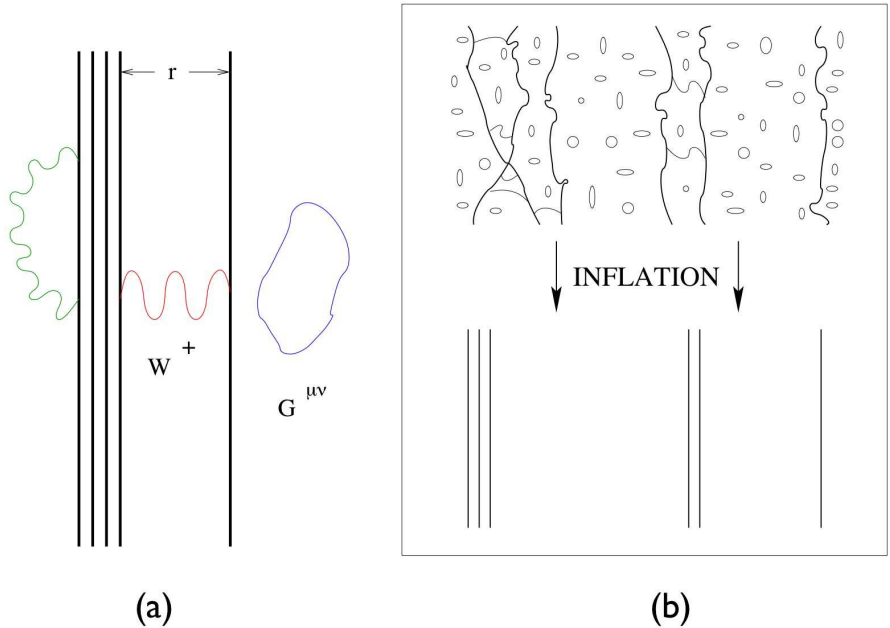} 
\vspace{0.1in}
\caption{ (a) The brane world scanrio. Here, as light open string modes with each end of an open string ending on a brane, the standard model particles are stuck to the branes, while closed string modes 
such as a graviton are free to roam the bulk. (b) During brane inflation, a tiny region of the branes (i.e., our universes) grows by an exponentially large factor. Fluctuations such as defects, radiation or matter will be inflated away. Also, the differences in spacing between branes as well as the curvature decreases rapidly.
}
\label{1.1}
\end{center}
\end{figure}

Since the discovery of D-branes in string theory \cite{Polchinski:1995mt}, a natural realization of nature in string theory is the brane world. In the brane world, all standard model particles are open string modes. Since each end of an open string must end on a brane, the standard model particles (being light) are stuck on a stack of $Dp$-brane, where 3 of the $p$ dimensions span our universe of standard model particles, while the remaining $p-$3 dimensions are wrapping some cycles in the bulk (the remaining $9-p$ spatial dimensions) where closed string modes such as the graviton live (see Figure \ref{1.1}(a)).
Suppose our today's universe is described by such a brane world solution in string theory.
A simple, realistic and well-motivated inflationary model is the brane inflation, where the inflaton is simply the position a $Dp$-brane moving in the bulk  \cite{Dvali:1998pa}. In the simple $D$3-${\D}$3-brane inflation \cite{collection}, inflation takes place while the $D$3-brane is moving towards the ${\D}$3-brane (i.e., anti-$D$3-brane, which has the same tension but opposite RR charge as a $D$3-brane) inside the 6-dimensional bulk (due to the attractive force between them), and inflation ends when they collide and annihilate each other. Fluctuations that are present before inflation, such as defects, radiation or matter, will be inflated away (see Figure \ref{1.1}(b)). Here, the relative $D$3-${\D}$3-brane position $\phi$ is the inflaton and the inflaton potential $V(\phi)$ comes from their tensions and interactions. The annihilation releases the brane tension energy that heats up the universe to start the hot big bang epoch. Typically, strings of all sizes and types may be produced during the collision. Large fundamental strings and/or $D$1-branes (or D-strings) that survive the cosmological evolution become cosmic superstrings.

\begin{figure}
\begin{center}
\includegraphics[width=0.8\textwidth,angle=0]{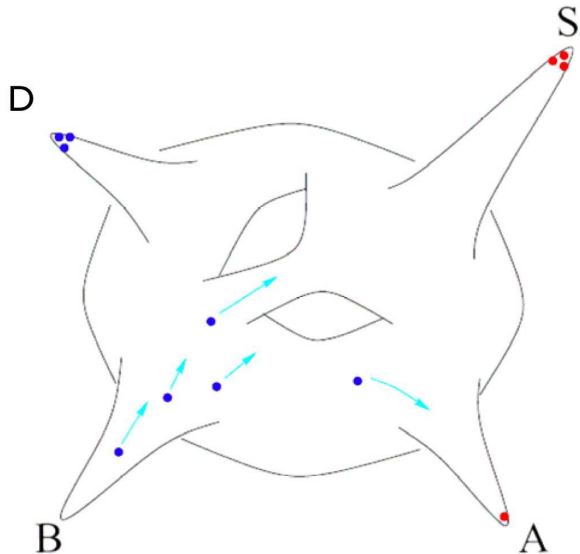} 
\vspace{0.1in}
\caption{A pictorial sketch of the compactified bulk. Besides some warped throats, there are $D$7-branes wrapping 4-cycles. The $D$3-${\D}$3-brane inflationary scenario in a generic flux compactfied 6-dimensional bulk.  The blue dots stand for mobile $D$3-branes while the red dots are ${\D}$3-branes sitting at the bottoms of throats. After inflation and the annihilation of the last $D$3-brane with the ${\D}$3-brane in $A$-throat, the remaining ${\D}$3-branes in $S$-throat may be the standard model branes. 
}
\label{1.2}
\end{center}
\end{figure}

In a more realistic brane world scenario, all moduli of the 6 extra spatial dimensions are dynamically stabilized via flux compactification \cite{Giddings:2001yu,Kachru:2003aw}, and the presence of RR fluxes introduces intrinsic torsion and warped geometry, so there are regions in the bulk with warped throats (Figure  \ref{1.2}). They are 6-dimensional versions of the Randall-Sundrum warped geometry. There are numerous such solutions in string theory, some with a small positive vacuum energy (cosmological constant). This is known as the string landscape. Presumably the standard model particles are open string modes; they can live either on $D$7-branes wrapping a 4-cycle in the bulk or (anti-)$D$3-branes at the bottom of a warped throat (Figure \ref{1.2}). In the early universe, there is an extra pair of $D$3-${\D}$3-branes.
Due to the attractive forces present, the ${\D}$3-brane is expected to sit at the bottom of a throat. Here again, inflation takes place as the $D$3-brane moves down the throat towards the ${\D}$3-brane, and inflation ends when they collide and annihilate each other, allowing the universe to settle down to the string vacuum state that describes our universe today. This is the KKLMMT scenario \cite{Kachru:2003sx}.
Although the original toy model version encounters some fine-tuning problems, the scenario becomes substantially better as we make it more realistic : it is surprisingly robust, that is, many e-folds of inflation is a generic feature. 
This is very encouraging. Briefly speaking, there are 2 key stringy ingredients that come into play: \\
$\bullet$ Because of the warped geometry, a consequence of flux compactification, a mass $M$ in the bulk becomes $h_{A}M$ at the bottom of a warped throat, where $h_{A} \ll 1$ is the warped factor (Figure \ref{1.2}). This warped geometry tends to flatten, by orders of magnitude, the inflaton potential $V(\phi)$, so
the attractive $D$3-${\D}$3-brane potential is rendered exponentially weak in the warped throat. 
The potential takes the form 
\ba
\label{infpot}
V(\phi) = V_K + V_A + V_{D \bar D} = \frac{1}{2}\beta H^2 \phi^2  
+ 2T_3h_A^4(1-\frac{1}{N_A}\frac{\phi_A^4}{ \phi^4}) + ...
\ea
where the first term $V_K(\phi) = m^{2}\phi^{2}/2 + . . . . $ receives contributions from the K\"{a}hler potential and various interactions in the superpotential \cite{Kachru:2003sx} as well as possible D-terms
\cite{Burgess:2003ic}. $H$ is the (initial) Hubble parameter so this interaction term behaves like a conformal coupling.  Here, $\beta$, and more generally $V_K$, probes the structure of the flux compactification \cite{Berg:2004sj,Baumann:2006th}. The warp factor depends on the details of the throat. Crudely, $h(\phi) \sim \phi/\phi_{edge}$, where $\phi=\phi_{edge}$ when the $D$3-brane is at the edge of the throat, so $h(\phi_{edge}) \simeq 1$. At the bottom of the throat, where $\phi=\phi_{A}$, $h_{A} = h(\phi_{A})= \phi_{A}/\phi_{edge}$. 
$T_{3}$ is the $D$3-brane tension and the effective tension is warped to a very small value $T_{3}h_A^4$ (as we shall see, $h_{A} \sim 10^{-2}$). The attractive gravitational (plus RR) potential is further warped to a very small value : $N_{A} \gg 1$ is the $D$3 charge of the throat.  If the last 55 e-folds of inflation takes place inside the throat, then $\phi_{edge} \ge \phi \ge \phi_{A}$ during this period of inflation. 
Note that $\beta$ is expected to be of order unity, $\beta \sim 1$. Despite the warped geometry effect, the above potential yields enough inflation only if $\beta$ is small enough, $\beta \lesssim 1/5$ \cite{Firouzjahi:2005dh}. 
\begin{figure}
\begin{center}
\includegraphics[width=0.8\textwidth,angle=0]{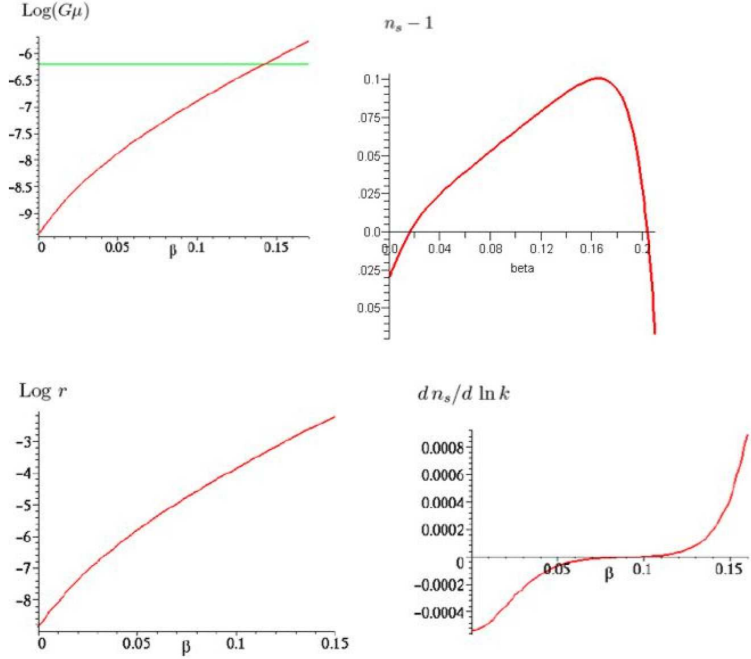} 
\vspace{0.1in}
\caption{The predictions of the slow-roll brane inflationary scenario \cite{Kachru:2003sx,Firouzjahi:2005dh} : the cosmic string tension $\mu$,
the power spectrum index $n_{s}$, the ratio $r$ of the tensor to the scalar density perturbations and the running of $n_{s}$. 
}
\label{1.3}
\end{center}
\end{figure}
We see in Figure \ref{1.3} that the data can easily over-constrain the model. However, this is not the end of the story. \\
$\bullet$ Because the inflaton is an open string mode, its kinetic term appears inside the Dirac-Born-Infeld action.
For slow-roll, this term reduces to the usual kinetic term. However, when the inflaton is moving relativistically, the full effect of the DBI action must be taken into account  \cite{Silverstein:2003hf}. 
The DBI action in brane inflation leads to the ``Lorentz factor''
\be
\label{gamma1}
\gamma(\phi)=\frac{1}{\sqrt{1-{\dot{\phi}^2}/T(\phi)}}
\ee
where $T(\phi)={T_{3}h(\phi)^{4}}$ is the warped $D$3-brane tension and  the limiting speed, $c(\phi)=\sqrt{T(\phi)}$, is decreasing rapidly as the $D$3-brane moves down the throat $c \sim \phi^{2} \rightarrow \phi_{A}^{2}$. This means the speed $\dot{\phi}$ 
of $\phi$ is limited by the rapidly decreasing limiting speed irrespective of the steepness of the inflaton potential. In the warped throat, even for a steep potential,
the inflaton motion must slow down considerably towards the bottom of the throat as it is becoming ultra-relativistic, so it takes a while before it reaches the bottom of the throat. 

As a result, the warped geometry of the throat combined with the DBI action generically allows for many e-folds of inflation. Robustness of the overall scenario suggests that we are in the right direction. 
A few comments are in order here :  \\
(i)  Since the inflaton is an open string mode that stretches between the branes, it no longer exists as a physical degree of freedom after the $D$3-${\D}$3-brane annihilation.\\
(ii) The above scenario does not guarantee enough inflation; however, it does yield enough inflation for 
a large region in the parameter space. Once CMBR and other cosmological data are introduced, constraints on the parameters will sharpen the predictions. At the moment, data is already putting strong constraints on the parameter space. Future data will constrain the parameters further and tell us about the structure of the bulk as well as the throat.\\
(iii) The presence of a $D$3-$\D$3-brane pair explicitly breaks supersymmetry. Although this breaking is large, but it is very soft, as we shall see. Furthermore, the warping exponentially suppresses the breaking terms. So it is justified to study the scenario within the supergravity approximation when the string scale is much smaller than the Planck scale. \\
(iv) The interplay between cosmology and gauge/gravity duality should receive more attention, since cosmological data may provide valuable information about strongly coupled gauge theory (via structures of throats and cosmic string properties). \\
(v) There are many variations of the above scenario. 
For large $m$ \cite{Alishahiha:2004eh}, or for a modified warped throat \cite{Dymarsky:2005xt}, enough inflation can be obtained without the $\D$3-brane. 
$D$3-$\D$3-brane inflation can take place in the bulk. 
Since a typical compactification has $D$7-branes wrapping 4-cycles in the bulk, one can also consider $D$3-$D$7-brane inflation \cite{Dasgupta:2002ew}. 
Multi-throat and/or multi-brane scenarios are also very easy to envision \cite{Iizuka:2004ct}. 
A crowd of $D$3-branes is quite natural, as illustrated in Figure \ref{1.2}, where many $D$3-branes can be released during a brane-flux annihilation in the $B$-throat \cite{Kachru:2002gs}. 
A nice scenario is the multi-throat model, where inflation takes place as $D$3-branes are moving out of the $B$-throat \cite{Chen:2004gc}.
There is  a large set of multi-brane inflationary models under the name ``assisted inflation'' \cite{Liddle:1998jc}. Clearly they should be fully explored. Some of the phenomenology is not too
 different from N-inflation \cite{Dimopoulos:2005ac}. Another variation is Ref.\cite{Huang:2006zu}. \\
(vi) The 6-dimensional (or 7-dimensional in M theory) compactification typically introduces many light closed string modes known as moduli. They include Kahler moduli, complex structure moduli, the dilaton
and KK type RR fields. The resulting effective potential involving these bulk modes is in general complicated enough so, with some fine-tuning, one can find a flat enough direction to carry out inflation. It is entirely possible that nature takes this path and moduli inflation should be and has been extensively studied. However, the moduli inflationary scenario does not seem to have distinct stringy signatures, or as compelling and predictive as brane inflation. \\

The rest of this paper discusses the various aspects of the above scenario:  \\
$\bullet$ Inflation. For small $m$ or $\beta$, the model reduces to the slow-roll scenario (Figure \ref{1.3}). For any given $\beta$, we see that the power spectrum index $n_{s}$, the ratio $r$ of the tensor to the scalar density perturbations, the running of $n_{s}$ and the cosmic string tension $\mu$ are determined \cite{Firouzjahi:2005dh}. In this case, WMAP and other cosmological data imposes the constraint $\beta <0.05$ \cite{Seljak:2006hi}. That is, the range $0.05 \lesssim \beta \lesssim 0.2$ is ruled out. 

For large inflaton mass $m$, the DBI action and so the warp factor comes into play. In this case, new stringy features such as non-Gaussianity will appear \cite{Alishahiha:2004eh}. Furthermore, the 3-point correlation function (or bispectrum) has a distinct distribution that is clearly different from what may appear in a slow-roll scenario \cite{Chen:2006nt,Maldacena:2002vr}. For intermediate values of $m$, the tensor mode perturbation may be large \cite{Shandera:2006ax}. It can also be distinguished from that coming from the slow-roll scenario. This is encouraging since, unlike the scalar mode perturbation, the metric perturbation directly probes the very early universe. \\
$\bullet$ Heating at the end of inflation. The $D$3-${\D}$3-brane annihilation produces only closed strings, with the graviton as the lightest mode. The transfer of energy from closed string modes to the  standard model particles which are open string modes seems problematic, since gravitational radiation can make up at most a few percent of the density of the standard model particles during big bang nucleosynthesis. Naively, this problem seems most severe if inflation takes place in one throat (the 
$A$-throat) while the standard model branes are in another throat (the $S$-throat). It is satisfying that an analysis of what happens indictates that heating will work out nicely. In fact, the situation improves dramatically when one considers a realistic (i.e., flux compactification) scenario instead of a toy model version based on the Randall-Sundrum scenario. It also offers some possibilities of specific features 
(such as KK modes as hidden dark matter \cite{Chen:2006ni}) that may be tested. \\
$\bullet$ Production and properties of cosmic strings. \\
$\bullet$ Evolution of the cosmic string network and its possible detection.
Here, we discuss our present knowledge of the scaling cosmic string network and some of its observational consequences.  

The history of cosmic strings is a long one \cite{kolb,Vilenkin}. First proposed by Kibble and others,
it was applied to generate density perturbations that seeded the structure formation.
This requires a tension of $G \mu \sim 10^{-6}$. This was ruled out by the CMBR data.
The possibility of superstrings as cosmic strings was first studied by Witten \cite{witten}. 
However, in the heteroric string framework, $G \mu \sim 10^{-3}$, which is far too big to be 
compatible with observations. 
In any case, either these cosmic strings would have been inflated away, or they are unstable to breakage. 
In brane inflation in Type IIB theory, we see that they are produced after 
inflation\cite{cosmicstring,Dvali:2003zj}, with much lower tensions due to the warped 
geometry \cite{Kachru:2003sx,Firouzjahi:2005dh}. 
They are stable or meta-stable enough (i.e., lifetime comparable to the age of the universe) under a variety of situations \cite{Copeland:2003bj,Leblond:2004uc,Polchinski:2005bg}, so they can survive to form a scaling cosmic string network. 
Cosmic superstrings will also have non-trivial tension spectrum and junctions can appear \cite{Copeland:2003bj}.
Of course, the presence of a cosmic string network is not guaranteed. However, if they 
are around, the chances of detecting them are very promising. 
Irrespective of the final outcome, we see that string theory is confronting data and 
making quantitative as well as distinctively stringy predictions. 

\section{Brane Inflation}

It is possible (in fact one may argue likely) that the inflaton potential has relatively flat directions outside the throat, allowing substantial inflation \cite{Hsu:2003cy}. Unfortunately, the precise potential is rather dependent on the detailed structures of the compactification and remains to be explored more carefully. To avoid this issue, we shall assume here that the $D$3-brane starts close to or inside the throat. If we have enough e-folds in the throat, then the physics outside the throat need not concern us. As explained earlier, this is an easy condition to satisfy.

\begin{figure}
\begin{center}
\includegraphics[width=0.8\textwidth,angle=0]{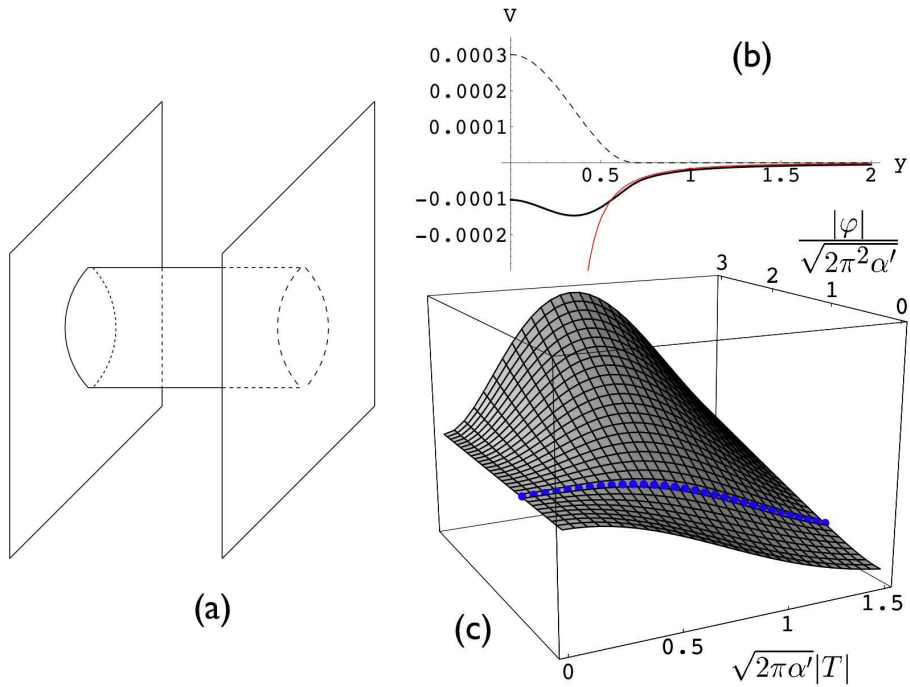}
\vspace{0.1in}
\caption{(a) The exchange of closed strings between two branes. In the dual channel, this describes  the one-loop radiative effect of the open strings stretching between 2 branes. (b) The potential $V(y)$ between the $D$3-brane and the ${\D}$3-brane due to the diagram (a), as a function of the separation $y$ for the brane pair, where $\ap=$1 \cite{Sarangi:2003sg}. 
The dashed curve is the imaginary part of $V(y)$. 
The thick line is the real part of $V(y)$. The Coulombic 
potential (the thin red curve) is shown for comparison. (c) The potential  $V(\phi, T)$ as a function of the inflaton $y \sim \phi$ and the tachyon expectation value $T$ \cite{Jones:2002si}. Brane inflation is a hybrid inflationary scenario.
}
\label{2.1}
\end{center}
\end{figure}

First, let us consider the potential $V(y)$ per unit volume
between a parallel $Dp-\D p$-brane
pair separated by a distance $y$, where the $Dp$-branes are BPS with 
respect to each other. We shall consider $p < 7$, where $T_p$ is the $Dp$-brane tension. 
We may view $V(y)$ as coming from the closed string exchanges between the branes (Figure \ref{2.1}(a)).
In the closed string channel, at large $y$, when the massive mode exchanges are Yukawa-suppressed,
\be
V(y) \simeq - \frac{\kappa^2T_p^2}
{ \pi^{(9-p)/2}}
\Gamma((7-p)/2) \frac{1}{y^{7-p}} 
\ee
where $\kappa^2=8 \pi G_{10}$ and $T_{p}= (2\pi \ap)^{-(p+1)/2}$ is the $D$p-brane tension.
Here $\ap=m_{s}^{-2}$ is the Regge slope and $m_{s}$ is the string scale.
For $p<7$, $V(y)$ vanishes as $y \rightarrow \infty$.
This is simply the attractive gravitational (NS-NS) plus massless RR interaction
between the branes.
At short distances, the exchange of the massive closed string modes are not Yukawa-suppressed and the evaluation of $V(y)$ is somewhat subtle. Because of the exponentially growing degeneracy (as a function of mass) in the closed string spectrum, a naive summation yields an oscillating divergent result.
Looking at Figure \ref{2.1}(a), we see that we may evaluate $V(y)$ as a 1-loop radiative correction in the open string channel by including the whole tower 
of open string modes. The particular way of grouping the contributions should be dictated by the 
soft supersymmetry breaking \cite{Garcia-Bellido:2001ky,Sarangi:2003sg}.
When the two branes are parallel there is no potential between 
them because of supersymmetry. Each mass level contains a 
set of supermultiplets. The contribution to the potential 
$V(y)$ from the open string bosons is exactly cancelled by the 
contribution from the open string fermions, mass level by mass level.
Now we consider the $\D p$-brane as a $Dp$-brane rotated by $\pi$.
Supersymmetry broken by this rotation is large, in the sense that level crossings take place. 
However, the supersymmetry breaking is very soft, that is, the open string spectrum follows
the spectral flow. For each broken supermultiplet, 
\ba
\sum_{i}(-1)^{F}m_{i}^{2n}=0, \quad \quad n=1,2,3
\ea
where $i$ runs over the spectrum in each large but ``softly broken'' supermultiplet (and $F$ is the fermion number). 
Keeping this grouping in the sum over the open string spectrum yields a finite $V(y)$
 (Figure \ref{2.1}(b)). 
This very soft SUSY breaking also justifies the continuous use of the supergravity formulation.

In the open string one-loop channel, a tachyon appears at short distances,
\be
\label{tachyonmass}
\ap m_{tachyon}^2 =  \frac{y^2}{4 \pi^2 \ap}  - \frac{1}{2}
\ee 
which contributes an imaginary part to $V(y)$. We see that the Coulombic form is a very good approximation before the tachyon appears, by which time inflation is over anyway. 
With $\phi=\sqrt{T_{3}}y$, the tachyon appears when $\phi=\phi_{E}$, and the annihilation process begins. Note that the potential flattens around this distance and the usual sharp drop in the Coulombic potential is absent.  
The potential $V(\phi, T)$ in Figure \ref{2.1}(c) is evaluated using boundary superstring field theory method \cite{Jones:2002si}.
So we have $ \phi_{i}>\phi_{55}>\phi_{E}>\phi_{A}$, where $\phi_{i}$ is the initial
$D$3-brane position when inflation starts and $\phi_{55}$ is the value of $\phi$ at 55 e-folds before inflation ends. So the scenario is a hybrid inflation. 
In the more realistic KKLMMT scenario, $V(\phi)$ becomes $V_{D \bar D}(\phi)$ given in Eq.(\ref{infpot}).

Warped throats such as the Klebanov-Strassler (KS) warped deformed conifold \cite{Klebanov:2000hb} are generic in any flux compactification that stabilizes the moduli. The Dirac-Born-Infeld (DBI) action for the inflaton field follows simply because the inflaton is an open string mode. 
By now it is clear that enough inflation is generic in this scenario thanks to : 
(1) the warped geometry of the throat in a realistic string compactification, which tends to flatten (by orders of magnitude) the attractive Coulombic potential between the $D$3-brane and the $\D$3-brane \cite{Kachru:2003sx}; The warped geometry also reduces the vacuum energy that breaks suersymmetry,
so the supergravity approximation is expected to be valid.
(2) The warped geometry of the throat combined with the DBI action, which forces the inflaton to move slowly as it falls towards the bottom of the throat, as pointed out by Silverstein and Tong \cite{Silverstein:2003hf}. In fact, one may get enough e-folds just from around the bottom of the throat \cite{Kecskemeti:2006cg}.

Inside the throat, the metric takes the form
\be
ds^2=h^{2}(r)(-dt^{2} + a(t)^{2}dx^2) + h^{-2}(r)(dr^2+r^2ds_{5}^2),
\label{10dmetric}
\ee
and the potential takes the simple approximate form (\ref{infpot}),
\be
\label{potential}
V(\phi) = V_K(\phi) + V_0 +V_{D \bar D}(\phi) \simeq \frac{m^2}{2}\phi^2 + V_{0}\left(1-\frac{V_{0}}{4\pi^{2}v}\frac{1}{\phi^4}\right) 
\ee
where the constant term $V_{0}=2T_3h_A^4= 2 T_{3} h(\phi_{A})^{4}$
is the effective vacuum energy.
The factor $v$ depends on the properties of the warped throat, with $v=16/27$ for the KS throat.
With some warping (say, $h_{A} \simeq 1/5$ to $10^{-3}$), the attractive Coulombic potential 
$V_C(\phi)$ can be very weak (i.e., flat).
The quadratic term $V_K(\phi)$ receives contributions from a number of sources and is rather model-dependent. However $m^{2}$ is expected to be comparable to $H_{0}^2=V_{0}/3M_{p}^{2}$, where $M_{p}$ is the reduced Planck mass ($G^{-1}=8 \pi M_{p}^{2}$). This sets the canonical value for the inflaton mass $m_{0}=H_{0}$ (which turns out to be around $10^{-7}M_{p}$). 

The scale of the throat $R$ is given by \cite{Gubser:1998vd}
\be
\label{throatR}
R^4 = 4\pi g_sN_A \alpha^{\prime2}\frac{\pi^{3}}{V(s_{5})}
\ee 
where $V(s_{5})$ is the $s_{5}$ volume. For the KS-throat, $V(s_{5})= v \pi^{3} =16\pi^{3}/27$.
For a generic value of $m$, usual slow-roll inflation will not yield enough e-folds of inflation. Ref.\cite{Firouzjahi:2005dh} shows that $m \lesssim m_{0}/3$ will be needed. Na\"{i}vely, a substantially larger $m$ will be disastrous, since the inflaton will roll fast, resulting in very few e-folds in this case. However, for a fast roll inflaton, string theory dictates that we must include higher powers of the time derivative of $\phi$, in the form of the DBI action
\be
S=-\int d^4x\;a^3(t)\left[T\sqrt{1- \dot{\phi}^2/T} + V(\phi) - T \right]
\ee
where $T(\phi) = T_{3}h(\phi)^{4}$ is the warped $D$3-brane tension at $\phi$. For the usual slow-roll,
$T\sqrt{1- \dot{\phi}^2/T}  - T \simeq  \dot{\phi}^2/2$, reproducing the standard kinetic term.
It is quite amazing that the DBI action now allows enough e-folds even when the inflaton potential is steep \cite{Silverstein:2003hf,Alishahiha:2004eh}. As the $D$3-brane approaches $\D$3-brane, $\phi$ and $T(\phi)$ decrease, and $h(\phi) \rightarrow h(\phi_{A})$. 
The key is that $\dot{\phi}$ is bounded by the limiting speed, and this bound gets tighter as $T(\phi)$ decreases. This happens even if the potential is steep, for example, when $m > H_{0}$. 
So the inflaton rolls slowly either because the potential is relatively flat (so $\gamma \simeq1$ in the usual slow-roll case), or because the warped tension $T(\phi)$ is small (so $1 \ll \gamma < \infty $).  
As a result, it can take many e-folds for $\phi$ to reach the bottom of the throat. 
When $\gamma \gg 1$, the kinetic energy is enhanced by a Lorentz factor of $\gamma$. Note that the inflaton is actually moving slowly down the throat even in the ultra-relativistic limit.
However, the characteristics of this scenario are very different from the usual slow-roll limit, where $\gamma \simeq 1$.
To draw a distinction, we call this the ultra-relativistic regime. 

\begin{figure}
\begin{center}
\includegraphics[width=0.8\textwidth,angle=0]{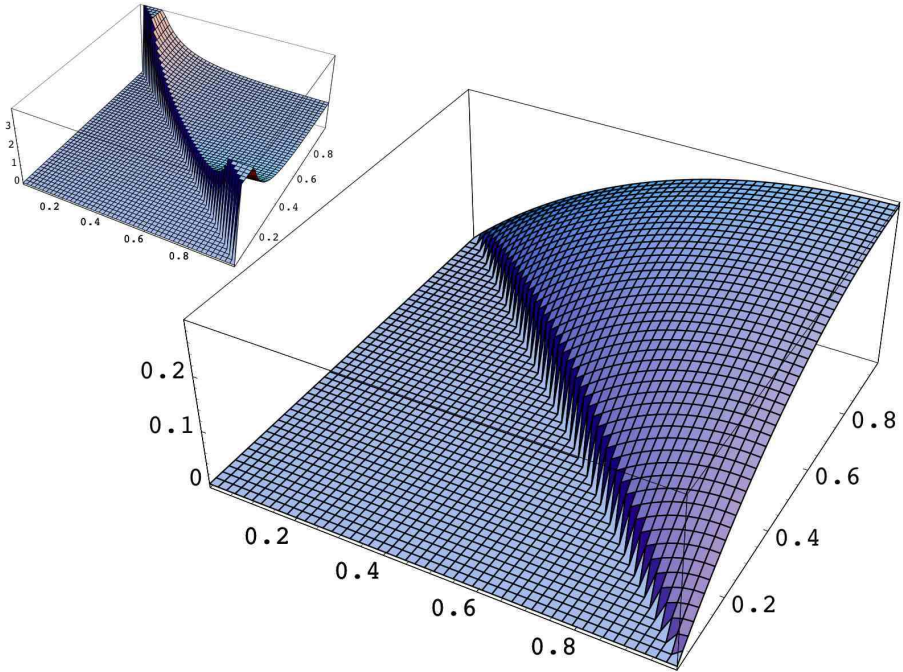}
\vspace{0.1in}
\caption{The shape of the 3-point correlation function in the DBI model \cite{Chen:2006nt}. For comparison, the (negative) of the 3-point correlation function from a standard slow-roll model is 
shown at the upper left corner. }
\label{2.3}
\end{center}
\end{figure}

In general, there are 3 parameters, namely, $m$, $\lambda$ and $\phi_{A}$ (note that $V_{0}$ is a function of $\lambda$ and $\phi_{A}$), plus the constraint that the $D$3-brane should be inside the throat. We find that the power spectrum can be red-tilted in all three scenarios. \\ 
(1) $\beta \ll 1$, $\gamma \simeq 1$, the slow-roll case, when $m^{2} \simeq 0$; 
Here, there are essentially 2 parameters : $m$ and $V_{0}$. After fitting the COBE density perturbation data  \cite{Smoot:1992td}, the predictions are reduced to a one-parameter, namely $\beta$, analysis \cite{Firouzjahi:2005dh}. For small $\beta$, $n_{s} \sim 0.98 +\beta$, $\log r \sim -8.8 +60\beta$, $\log G\mu \sim -9.4 + 30 \beta$. The cosmological data restricts the relevant range to $0 \le \beta < 0.05$
\cite{Seljak:2006hi}.\\
(2) $\beta \sim 1$, $\gamma \simeq 1$ at $N_{e} \sim 55$, but increases to a large value towards the end of inflation; this corresponds to some intermediate values of $m^{2}$. Here, the DBI introduces a deviation from the slow-roll relation between $R$ and the tensor power spectrum index $n_{t}$  \cite{Shandera:2006ax} , 
\be
n_t=-\frac{r}{8}\left(\frac{\gamma}{1-\epsilon -\kappa}\right)
\ee
where $\epsilon$ is the usual slow-roll parameter divided by $\gamma$ and $\kappa$ measures the running of $\gamma$. For large $\phi$, the parameterization of the potential should probably include a $\phi^{4}$ term. Naively, the tensor mode can be large, i.e., as large as saturating the present observational bound $r <0.3$ \cite{Spergel:2006hy}. However, the position of the $D$3-brane is bounded by the bulk volume. Applying the Lyth bound (unmodified by the DBI action): 
$ \Delta \phi /M_{P} =\sqrt{r/8} \Delta N_{e}$, one sees that, to get enough e-folds while keeping $\phi$ inside the compactified volume, $r$ cannot be bigger than a few percent \cite{Lidsey:2006ia}. Such a large value of $r$ would require $r$ to drop rapidly as the inflaton moves down the throat. This can happen if $\gamma$ increases rapidly \cite{Shandera:2006ax}. 
This constraint on $r$ is somewhat relaxed if we have a crowd of $D$3-branes.
A large $r$ may also indicate that the warped throat is somehow squashed. We note that, even with a DBI action, the tachyon rolling by itself can yield at most a few e-folds \cite{Leblond:2006cc}.  \\
(3) $\beta \gg 1$, $\gamma$ is large throughout. In this ultra-relativistic case, $m$ is large so $V_{K}$ dominates (i.e., $V_{0}$ can be ignored), and the model is again reduced to the above 3 parameters before imposing the COBE normalization. In this scenario, ensuring that all 55 e-folds of inflation take place while the $D$3-brane is inside the throat becomes a strong constraint; that is, the ``initial'' position $\phi_{i}$ (at 55 e-folds before the end of inflation) should satisfy  $\phi_{i} \le \phi_{e}$ where $\phi_{e}$ is the value at the edge of the throat, i.e., $h(\phi_{e}) \simeq 1$. To implement this condition, we need to introduce the $D$3-brane tension $T_{3}$, or the string scale $\ap$.
Since $V_{0}$ can be ignored in this case, one may obtain all the inflationary properties without the $\D$3-brane. 
 $|f_{NL}|  \simeq 0.32 \gamma^{2} \lesssim 300$ yields $\gamma \lesssim 31$ \cite{Creminelli:2005hu}.
However, one should check if reheating or preheating can be successfully realized in such a scenario. 
The structure of the non-Gaussianity from this UV DBI model is different from that due to slow-roll. The 3-point correlation functions (bi-spectrum) $A(k_{1}, k_{2})/k_{1}k_{2}k_{3}$ 
(where $k_{1}+k_{2}+k_{3}=0$) \cite{Chen:2006nt,Maldacena:2002vr} are shown in Figure \ref{2.3}. The 4-point correlation functions (tri-spectrum) have also been studied \cite{Huang:2006eh}.

Note that $n_{s}$ is quite sensitive to the warped factor. This point is clearly illustrated by the 2 different predictions of $n_{s}$ using 2 different approximations to the KS warp factor: an AdS cut-off (very slightly blue tilt) \cite{Shandera:2006ax} and a mass gap cut-off (red tilt) \cite{Kecskemeti:2006cg}. 
For large $R$, we have to consider a highly orbifolded version of the throat in order to fit it inside the bulk. This tends to suppress the DBI non-Gaussianity. \\
(4) For tachyonic inflaton mass ($m^{2} <0$), the scenario becomes the multi-throat brane inflation scenario proposed by Chen \cite{Chen:2004gc,Chen:2005ad}. The Coulombic term $V_{D \D}$ is negligible and inflation takes place as the $D$3-brane moves out of a throat (see Figure \ref{1.2}). For small tachyonic mass, this is simply a slow-roll model. This IR DBI inflation can happen when inflaton mass takes a generic value, $m \approx H$ ($\beta \approx 1$).
The distance the inflaton travels through during inflation, $\Delta \phi
\approx H R^2 \sqrt{T_3}$, is always sub-Planckian.
This model may be realized in a multi-throat compactification starting
with a number of antibranes settled down at the ends of various throats.
These antibranes are classically stable, but can annihilate against
the fluxes quantum mechanically \cite{Kachru:2002gs}. The end products
of such a phase transition is many $D$3-branes in, say, the $B$-throat,
which is sufficiently long (typically more than twice longer than the $A$-throat). 
The IR model predicts large non-Gaussianity with the same shape as in the UV
model. The difference is the running, $f_{NL} \approx 0.036 \beta^2 N_e^2$, that is,
$f_{NL}$ decreases with $k$, while $f_{NL}$ increases with $k$ in the UV model. 
The power spectrum index undergoes an interesting phase transition at a critical e-fold 
from red ($n_s-1 \approx -4/N_e$) at small scales to blue ($n_s -1  \sim 4/N_e$) 
at large scales \cite{Chen:2005ad,Chen:2005fe}. 
This transition is due to the Hagedorn phase when the red-shifted string scale drops
 below the Hubble constant. 
If such a transition falls into the observable range of CMBR, it predicts a
large running of $n_s$ around the transition point, i.e., a large negative $dn_s/d\ln k$. 
Outside of this transition region, $dn_s/d\ln k$ is un-observably small. 
                                
If the brane inflationary scenario is correct, it will provide a great probe to both the origin of our 
early universe as well as to the particular compactification in string theory, i.e., where we 
are in the cosmic landscape. For example, the inflaton is actually a 6-component field. 
So far, we have only considered the radial mode. When a 4-cycle is close to the $A$-throat, 
the symmetry of the throat ($S^{3} \times S^{2}$ for the KS case) would be broken by the 
4-cycle's position, shape and orientation, generating a richer inflaton potential \cite{Baumann:2006th}.
CMBR data actually imposes conditions on the the structure around the throat \cite{Burgess:2006cb}.
That is, we can learn about the flux compactification via brane inflation. 
This may also tell us whether eternal inflation is happening or not. 
Since, $\phi$ is bounded by the size of the bulk, eternal inflation is far from 
a given in brane inflation \cite{Chen:2006hs}. 

\section{Graceful Exit}

The crucial step that links the inflationary epoch to the hot big bang epoch
is the heating at the end of inflation. This is known as the graceful exit,
namely, how the inflationary energy can be efficiently
transferred to heat up the Standard Model particles, and be compatible
with the well-understood late-time cosmological evolution? 
This is the heating problem (also called the reheating or preheating
problem). To see why this is quite a non-trivial issue, we first
look at the end process of brane inflation.

In the above brane inflationary scenario, inflation ends when the $D$3-brane annihilates 
with the $\D$3-brane. Significant insights have been gained into such a process
 \cite{Sen:2002nu}. 
Tachyonic modes appear when the brane-antibrane distance approaches the string scale
and the annihilation process may be described by tachyon rolling
\cite{Shiu:2002xp,Cline:2002it}. 
(The decay width is signified by the imaginary part of the potential $V(\phi)$).
No matter whether there are adjacent extra branes surviving
such an annihilation (for example, a $D$3-brane colliding with a stack of $\D$3-branes), 
the initial end product is expected to be dominated by
non-relativistic heavy closed strings \cite{Lambert:2003zr,Chen:2003xq,Leblond:2005km}.
These will then go to lighter closed strings, light KK modes, gravitons and open
strings. We know from observations that, during big bang 
nucleosynthesis (BBN), the density of gravitons can be no more than
a few percent of the total energy density of the universe. The rest is
contributed by the Standard Model (SM) particles (mostly photons,
neutrinos and electrons), which are open strings attached to a stack
of SM (anti-)branes.
We also know that the density of any non-relativistic relics can be no
more than about 10 times that of the baryons.
Therefore the question becomes how the brane annihilation products,
originally dominated by the closed string degrees of freedom, can eventually
become the required light open string degrees of freedom living on the
SM branes, with a negligible graviton density and a non-lethal amount
of stable relics. This question is particularly sharp in the 
the multi-throat scenario, where the inflationary branes
annihilate in one throat ($A$-throat) while the SM branes are sitting in another 
throat (say, $S$-throat in Figure \ref{1.2}). Let us discuss this case
and then comment on the other cases.

A number of studies have been done to address this heating problem
\cite{Barnaby:2004gg,Kofman:2005yz,Chialva:2005zy,Frey:2005jk}.
The analysis is essentially based on the Randall-Sundrum (RS) warped geometry. 
An important observation is that, because the KK mode wave function
is peaked at the bottom of the throat, its interaction with
particles located there is much enhanced compared to that
with the graviton, whose wavefunction spreads throughout the bulk. 
Because of this, the graviton emission branching ratio during the brane decay and KK
evolution is suppressed by powers of warp factors \cite{Dimopoulos:2001ui}. 
In a realistic compactification, throats are typically
separated in the bulk, which tends to generate resonance effects in
the tunneling from one throat to another. We expect the compactification 
volume to be dominated by the bulk, with typical size $L \gg R$, another important ingredient in the 
success of the graceful exit.
Again, the realistic scenario of heating improves in a number of ways over the RS scenario.
The discussions below follows Ref.\cite{Chen:2006ni} and relies on Figure \ref{1.2}.

First, we note that that the cross sections for KK self-interaction and KK interactions with SM particles
in a throat with size $R$ and a warp factor $h$ goes like
\ba
\label{crosst}
\sigma \sim \left( \frac{L}{R}\right)^{6} \frac{1}{M_{P}^{2}h^{2}}
\ea 
This is much bigger than that for the graviton, where the corresponding $\sigma \sim M_{P}^{-2}$. Note that the factor $(L/R)^{6}$ comes from the 6-dimensional bulk.
Next, it is important to follow the thermal history of the KK modes as the universe expands.
Because of the above warped enhanced KK self-interactions, it is easy to see that the KK modes become non-relativistic before they decouple. Scattering and annihilation of KK modes reduce the KK density by orders of magnitude. So, instead of a tower of high-density non-relativistic, non-interacting 
KK modes, only the lightest few stable KK modes remain and their relic density is very much suppressed. Gravitino density is similarly suppressed. The qualitative picture of heating goes as follows. 

Massive closed string modes produced during the $D$3-$\D$3-brane annihilation
rapidly decays to light KK modes and gravitons.
Among the light KK modes in a throat are ones with conserved angular momenta, 
so they are quite stable against further decay, with typical mass of order $h_{A}/R$. Due to the self-interaction, the relic density in the non-relativistic KK modes is very much suppressed.
Due to the red-shift and the low tunneling rate, the universe enters a matter-dominated phase
with these KK modes, which then tunnel to the $S$-throat and other throats, if present. 
To ameliorate the hierarchy problem, we expect the $S$-throat to have a much smaller warped 
factor $h_{S} \ll h_{A}$. Generically, we expect the tunneling rate from $A$-throat to $S$-throat to be enhanced by the bulk resonance effect (for $R/L \lesssim h_{A}$) \cite{Firouzjahi:2005qs},
\ba
\Gamma_{A \rightarrow S} \sim h_{A}^{9}/R \gg h_{A}^{17}/R
\ea  
where the second rate is that for the case when there is no bulk resonance effect.
Once the KK modes reach the $S$ throat, they rapidly decays to open string modes
and heat up the universe, starting the hot big bang epoch. For a successful scenario,
(1) the matter dominated duration should be long enough to red-shift away the gravitational 
radiation as well as the gravitino density, but not so long as to over-cool the universe. 
This condition requires
$h_{A} \sim 10^{-1}$ to $10^{-3}$. It is very encouraging that these values are precisely those required 
to fit the CMBR data.
(2) the decay of KK modes in the $S$ throat should go to open string modes instead of 
to gravitational radiation. This is guaranteed because the coupling of KK modes to 
gravitons is dictated by the Newton's constant $G_{4} =8 \pi/M_{P}^{2}$, while their couplings 
to open strings modes, i.e, SM particles, is enhanced by the 
localization of both the KK modes and the SM branes in the throat, as shown in Eq.(\ref{crosst}).

It is interesting to point out some novel features in this heating scenario: \\
$\bullet$ There is a matter-dominated epoch between the end of inflation and the beginning of the hot big bang era. The cosmic scale factor can grow by a large factor ($10^{5}$ or more) during this epoch.
As a result, both the gravitational radiation and the gravitino density will be substantially suppressed. 
It will be interesting to study other cosmological consequences of such an epoch. \\
$\bullet$ There is a dynamical process that selects a long throat to be heated. This is
because the dense spectrum in a long throat makes the level
matching of the energy eigenstates, a necessary condition for
tunneling between throats, easier to satisfy. This may provide a
dynamical explanation of the selection of the RS type (i.e., with very large warping that solves the hierarchy problem) warp space as our Standard Model throat in the early universe. \\
$\bullet$  Although KK modes as dark matter have been considered in the literature, we see
the possibility of KK modes as hidden dark matter. These are almost stable KK modes in another throat 
(say, the $D$-throat in Figure \ref{1.2}), which interacts only via gravitons with SM particles.
This hidden dark matter has many unusual properties compared to the
usual dark matter candidates, e.g., it may tunnel to the $S$-throat and generate a cosmic ray 
that violates the GZK bound.

More generally, it is found that heating is not a problem unless both the brane annihilation and the SM branes are in the bulk. Since it is natural for inflation and so brane annihilation to happen inside a throat, we consider heating to be a solved problem in brane inflation.

\section{Production and Properties of Cosmic Superstrings 
}

Although the production of domain walls and monopoles at the grand unified 
(GUT) scale will over-close the universe by many orders of magnitude, cosmic strings
do not suffer from the same problem. This is a consequence of the intercommutation properties of strings, which leads to a scaling cosmic string network that tracks the radiation (matter) during the radiation- (matter)-dominated era. A key property of cosmic string is its tension $\mu$.
In fact, cosmic strings around the GUT scale, that is $G \mu \sim 10^{-6}$, was originally proposed as an alternative to inflation in generating density perturbation for structure formation \cite{Vilenkin}. 
However, the properties of CMBR data, in particular the acoustic peaks, ruled out this possibility.
It is this same data that strongly supports inflation.
In fact, all defects present before inflation would have been inflated away.
So we need to consider only defects that are produced after inflation.

The topological properties of defect formation in tachyon condensation
is well understood in superstring theory \cite{Sen1}.
The spontaneous symmetry breaking will support defects with even
codimension (i.e., $2k$), as classified by K theory. 
In particular, $D$3-$\D$3-brane annihilation yields
$D$1-branes and fundamental $F$1-strings, when the large massive ones appear as 
cosmic strings in our universe \cite{cosmicstring,Dvali:2003zj,Copeland:2003bj}. 
Qualitatively, it is easy to see how this takes place.
There is a $U(1)$ gauge theory associated with each brane, and the tachyon couples to one combination $U(1)_{-}$. This is simply the Abelian Higgs model in the field theory approximation.
Tachyon rolling results in spontaneous symmetry breaking and the resulting vortices are 
$D$1-strings. So they are cosmologically produced via the Kibble mechanism.
The other $U(1)_{+}$ becomes confining, and the resulting flux tubes become the fundamental 
closed strings \cite{Bergman:2000xf}. So cosmic strings are generically produced towards the end of brane inflation. 
It is quite amazing that string theory dictates that the dangerous domain walls and monopole-like 
defects are not produced. In the Type IIB theory that we are studying, there is simply no $D$0- 
or $D$2-branes.

 We find that the cosmic string tension $\mu$ depends on the specific scenario. It roughly satisfies $10^{-13}<G\mu<10^{-6}$ \cite{cosmicstring,Firouzjahi:2005dh,Shandera:2006ax}.
Fundamental string (F-string) tension in 10 dimensions defines the string scale $\ap$ via  $T_{F1}=1/2 \pi \ap$. In type IIB theory, there are branes including $D$1-branes, or $D$-strings, with tension
$T_{D1}=1/2 \pi \ap g_{s}$, where $g_{s}$ is the string coupling.  In light of all the progress coming from dualities in string theory, we now know that the $D$-strings and the $F$-strings should be considered on the same footing and a general string state in type IIB is the bound state of these two types of strings.
In 10 flat dimensions, supersymmetry dictates that the tension of the bound state of $p$ $F$-strings and $q$ $D$-strings is given by \cite{Schwarz:1995dk},
	\ba
	\label{flat}
	T_{{p,q}} = T_{F1} \sqrt{p^2  +\frac{q^2}{g_s^2}}\, .
	\label{pqtension10}
	\ea
This tension spectrum (for coprime $(p,q)$) allows junctions to be formed \cite{Copeland:2003bj}.  
Since the $D$3-$\D$3-brane annihilation most likely takes place at the bottom of a throat, that will be where the cosmic superstrings are. 
To be specific, we consider the KS throat \cite{Klebanov:2000hb}  whose properties are relatively well understood. On the gravity side, this is a warped deformed conifold. Inside the throat, the geometry is a shrinking $S^{2}$ fibered over a $S^{3}$.
The tensions of the bound state of $p$ F-strings and that of $q$ $D$-strings were individually computed for the KS throat  \cite{Gubser:2004qj}. 
The tension formula for the $(p,q)$ bound states is given by \cite{Firouzjahi:2006vp}
\ba
\label{finalanswer}
T_{p,q} \simeq  \frac{h_{A}^{2}}{2 \pi \ap} \sqrt{  \frac{q^2}{g_s^2} + 
(\frac{b M}{\pi})^2 \sin^2(\frac{\pi p}{M})},
\ea  
where $b=0.93$ is a number numerically close to one and  $M$ is the number of fractional D3-branes, that is, the units of 3-form RR flux $F_3$ through the $S^{3}$.
For $M \rightarrow \infty$ and $b=h_{A}=1$, it reduces to Eq.(\ref{pqtension10}).
Very interestingly, the $F$-strings are charged in $\mathbb{Z}_M$ and are non-BPS. 
The D-string on the other hand is charged in $\mathbb{Z}$ and is BPS with respect to each other.  
Because $p$ is $Z_{M}$-charged with non-zero binding energy, binding can take place even if $(p,q)$ are not  coprime. 
Since it is a convex function, i.e., $T_{p+p'} < T_{p} + T_{p'}$, the $p$-string will not decay into strings with smaller $p$. 
The interpretation of these strings in the gauge theory dual is known. The $F$-string is dual to a confining string between a quark and an anti-quark, while the $D$-string is dual to an axionic string.  
$M$ fundamental strings can terminate to a point-like baryon (with mass $\sim M^{3/2} h_{A}/\sqrt{\ap}$), irrespective of the number of D-strings around.
	
	Besides the above Kibble and confining mechanisms, there are other possible ways to produce cosmic strings which may evolve to a cosmic string network : \\
$\bullet$ Consider another throat with warped factor $h_{C}$. If the temperature at the beginning of the hot big bang is $T_{i}$, then strings in the $C$-throat will be excited if $T_{i}>h_{C}m_{s}$. \\
$\bullet$ D-strings can be stable inside $D$3-branes \cite{Leblond:2004uc}. Such D-strings can be pair-produced inside the horizon at the end of inflation when a small stack of $D$3-branes collide with a larger stack of $\D$3-branes. \\
$\bullet$ One may also consider the situation when a single branes move towards the bottom of 
the $A$-throat. Assuming that heating is not a problem for such a scenario, stable D-strings might have been pair-produced if $T_{i} > m_{s}h_{A}$. 

Isolated loops would just decay via gravitational radiation. However, if the density of loops is high enough so that they overlap and tangle with each other, then 
their reconnections will yield bigger and bigger loops, which may eventually yield long strings and lead to a scaling cosmic string network.
For the $C$-throat, this probably requires $T_{i} \gg h_{C}m_{s}$. This is more likely for small $G \mu$, since the decay rate is proportional to $G \mu \sim G m_{s}^{2}h_{C}^{2}$, so light tension sting loops will be quite long lived.
This will imply that, in addition to cosmic strings in the $A$-throat, the universe may have cosmic strings with much smaller tensions if throats with large warping exist in the bulk. These light cosmic strings interact very weakly with cosmic strings in $A$-throat.

\section{Evolution and Detection of Cosmic Superstrings}

The cosmological evolution of cosmic superstrings is a very challenging problem. 
For slow-moving cosmic strings that stretch across the horizon, 
the energy density naively scale like $a^{-2}$. 
For cosmic string loops, the naive energy density is similar to that for 
monopoles, scaling like $a^{-3}$. So, naively, the cosmic string density is a problem.
However, their interactions substantially suppresses the density.
The intercommutation of intersecting cosmic strings and the decay of the 
resulting cosmic string loops (to gravitational waves) reduces the density so that it decrease like radiation (matter) during the radiation- (matter)-dominated era \cite{Vilenkin}.
Furthermore, the resulting scaling cosmic string network energy 
density is insensitive to the initial density, that is, the network rapidly approaches the scaling solution.
As a consequence, the physics is essentially dictated by the single parameter $G \mu$ in 
the Nambu-Goto or the Abelian Higgs model, and by the tension spectrum for a more complicated model.

Although the cosmic string network reaches a scaling solution,
the fraction of energy density in cosmic string loops has been an outstanding question
\cite{Vilenkin}. Early simulations did not reach fine enough resolution to determine the role 
played by string loops \cite{stringnet}. 
The basic assumption is that once a loop is produced by the intersection of long 
strings (including self intersection), they decay quickly via gravitational radiation.
More recent analysis seems to change the story. 

Let us first consider the Nambu-Goto case.
The fraction of energy density in the string network is given by
\be
\label{fdensity}
\Omega_{s} = \Omega_{\infty} + \Omega_{loops} \sim \Gamma G \mu + \chi \sqrt{\alpha G \mu}
\ee
where the first term is the contribution of long strings, with $\Gamma \sim 10^{2}$ for Nambu-Goto 
strings.
The second term is the contribution of string loops within the horizon. 
Very crudely speaking, $\chi \sim 10^{3}$. The value of
$\alpha$, the ratio of characteristic loop size to the horizon size, is poorly understood. It has been estimated to be as small as $\alpha < 10^{-12}$, or $ \alpha \sim G \mu$ or even $(G \mu ) ^{5/2}$.
Recently, both numerical simulations \cite{Vanchurin:2005yb,Avgoustidis:2005nv} and analytic studies \cite{Polchinski:2006ee} have indicated that there are more energies in the string loops than previously thought. That is, $\alpha$ may be as big as $0.25$, although $\alpha \sim 10^{-4}$ seems to be more likely. For small $G \mu$, the increase in the energy density in the string network can be very substantial.
\begin{figure}
\begin{center}
\includegraphics[width=0.8\textwidth,angle=0]{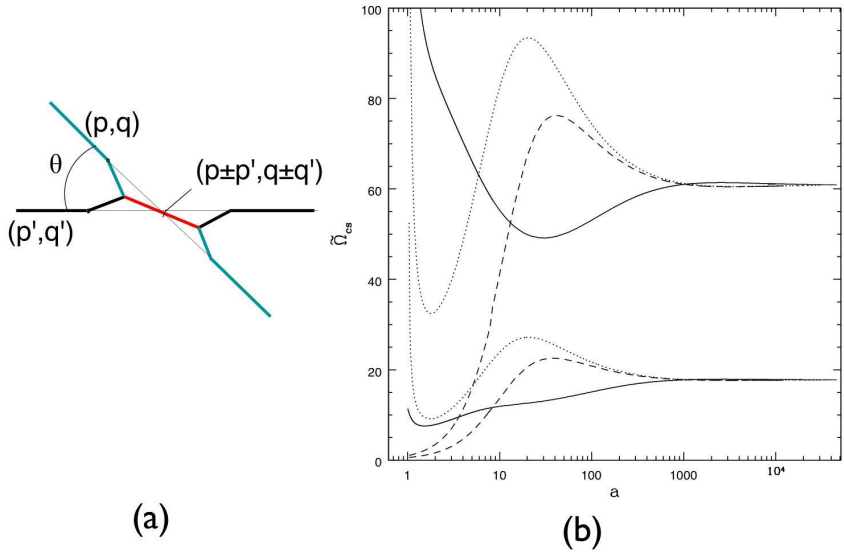}
\vspace{0.1in}
\caption{(a) The $(p,q)$ string binding generates junctions \cite{Copeland:2003bj}. (b) $(p,q)$ string network evolution as a function of the cosmic scale factor. The top 3 lines stand for total density with 3 different initial densities, while the bottom 3 lines stand for the corresponding $(p,q)=(1,0)$ string density with 3 different initial densities. We see that, irrespective of the initial densities,  both the total density and the $(1,0)$ density rapidly approach the scaling solutions \cite{Tye:2005fn}. 
}
\label{5.1}
\end{center}
\end{figure}

As mentioned earlier, cosmic superstrings will have different properties than vortices in the Abelian Higgs model. Although a simulation is not available, one can analyze the evolution of the string network by solving a set of coupled equations. As shown in Figure \ref{5.1}(b), recent analysis on the tension spectrum (\ref{flat}) strongly suggests that cosmic superstrings also evolve dynamically to 
a scaling solution (with a stable relative distribution of strings with different quantum numbers) 
\cite{Jackson:2004zg,Tye:2005fn}, very much like usual cosmic strings (either coming from the 
abelian Higgs model, or Nambu-Goto type) \cite{Vilenkin}. This is due to the rapid decrease in the density of strings with large tensions, which goes roughly like $\mu (p,q)^{-N}$, where $N \sim 8$.
We shall consider a scenario where the cosmic strings are stable enough to allow such a scaling solution.
The inter-commutation probability of vortices is known to be around unity, $P \simeq 1$,  while that of superstrings is rather complicated\cite{Jackson:2004zg}, but $P \sim g_{s}^{2}$, where the string coupling $g_{s} \sim 1/10$, is not unreasonable.  Also, the tension spectrum tells us that cosmic superstrings will come in a variety of tensions and charges. A simple analysis indicates that a number of species of cosmic strings will be around in the string network \cite{Tye:2005fn}, so a naive scaling from the Nambu-Goto strings to superstrings gives
$$\Omega_{s} \rightarrow \Omega_{superstring} \sim \frac{n}{P} \Omega_{s} =\frac{n}{g_{s}^{2}} \Omega_{s}$$
where $n$ is the effective number of types, $n \sim 5$. For very small $P$, it is argued that ${1}/{P} \rightarrow {1}/{P^{2/3}}$ \cite{Avgoustidis:2005nv}. This enhancement should help the search.
It is not clear how the presence of baryons in the tension spectrum (\ref{finalanswer})
will impact on the evolution of the string network. It is clear that further studies, 
the properties of cosmic string spectrum 
(including baryons), their productions, stabilities and interactions, and the cosmic evolution of the network as well as their possible detections will be most interesting to watch. It is reasonable to be optimistic about the detectability of cosmic superstrings, but this is far from guaranteed. 
Cosmologically, the strings appear to have beads, that is, they look like necklaces
This problem has been studied at some level \cite{Siemens:2000ty}. 

Originally proposed as an alternative to inflation, the detection of cosmic strings 
has been extensively studied \cite{Vilenkin}. Since the cosmic superstrings 
interact with the SM particles only via gravity, all detection involves the gravitational interactions of cosmic strings. Recent understanding on the importance of string loops will certainly enhance the detectability of cosmic strings. Since the particular brane inflationary scenario is not yet known, the cosmic string tensions are only loosely constrained. We shall be open-minded in comparing with observation. Many ways to detect cosmic strings have been suggested. 
Here, let us discuss some of them : \\
$\bullet$ Gravitational lensing is probably most direct. Cosmic string 
introduces a deficit angle, so a galaxy behind a long cosmic string
will appear as a double (undistorted) image. The image separation 
is roughly $5 \times 10^6 G \mu$ arc sec. For $G \mu \ll 10^{-7}$, this 
approach becomes very challenging.  
Finding a lensing by a junction will be quite definitive \cite{Copeland:2003bj,Shlaer:2005ry}. \\
$\bullet$ Micro-lensing. This was first studied in Ref.\cite{hogan}. 
For small string tension, string loops have relatively long lifetimes and so are expected 
to be dominant in the cosmic string energy density. 
They can lens stars, which shows up as the brightness of a star doubles for a short period of time.
Since there are more string loops for smaller tension, non-observation may put a lower bound on the cosmic string tension \cite{chernoff}. \\
$\bullet$ In brane inflation, the density perturbation (and CMBR anisotropy) comes from two 
sources:  the usual quantum fluctuation (scalar and tensor modes) during inflation and the 
fluctuations (scalar and vector modes) induced by the cosmic string network. 
The density perturbation coming from the cosmic string 
network is active and incoherent, so there is no acoustic peaks that are prominent 
in the density perturbation coming from inflation.
\begin{figure}
\begin{center}
\includegraphics[width=0.8\textwidth,angle=0]{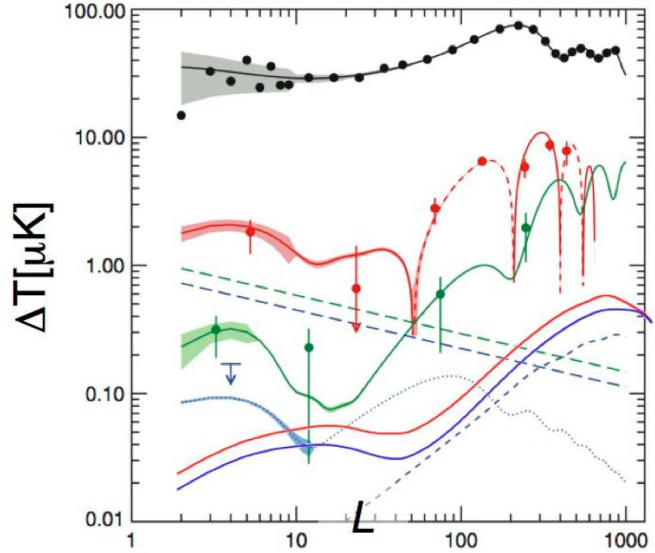}
\vspace{0.1in}
\caption{The CMBR power spectrum from WMAP \cite{Spergel:2006hy}. 
They are (from top) the temperature $TT$ correlation (black),
the temperature-electric-mode polarization $TE$ correlation (red), the $EE$ correlation (green), possible B mode polarization $BB$ correlation (blue) and possible $BB$ correlation (red/blue) from cosmic strings \cite{Pogosian:2003mz}. The dashed lines are likely background/foregound that should be subtracted.
}
\label{5.3}
\end{center}
\end{figure}
The COBE data roughly yields $G\mu \simeq 10^{-6}$ if the scaling 
solution of the cosmic string network is the sole source of the density 
perturbation. Using WMAP data, one finds that the contribution 
from cosmic strings is bounded by about 10 \%,
which translates to about $G\mu \lesssim 7 \times 10^{-7}$.
So the cosmic string production towards the end of brane inflation 
is perfectly compatible with the present CMBR data \cite{Spergel:2006hy}, while future data 
may be able to test this scenario \cite{smoot,Pogosian:2003mz}. \\
$\bullet$ Since the density perturbation coming from cosmic string is 
continuously being produced, its magnitude in the CMBR anisotropy
at large $l$ will not be attenuated as much as that coming 
from inflation. For $G\mu \sim 7 \times 10^{-7}$, the contribution 
from cosmic strings may become comparable to (bigger than) that from 
inflation at $l > 2000$ ($l > 3000$).
This may be measurable if $G \mu$ is not too small.
Polarization in the CMBR will also be measured. In particular, the
B (i.e., curl) mode due to the tensor mode perturbation will be tested,
reaching $\Delta T \simeq 0.5 \mu K$.
Here the gravitational wave anisotropy density is much higher than 
that in a pure inflationary scenario, so passage through
space will presumably yield a B mode polarization clearly larger 
than that coming from a purely inflationary 
scenario \cite{Pogosian:2003mz}. Figure \ref{5.3} illustrates this possibility. \\
\begin{figure}
\begin{center}
\includegraphics[width=0.9\textwidth,angle=0]{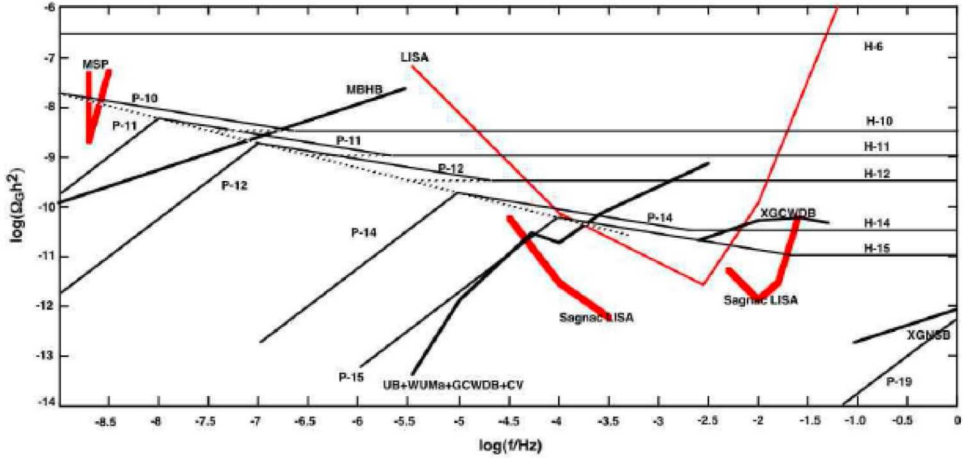}
\vspace{0.1in}
\caption{ 
The detectability of cosmic strings by LISA via gravitational radiation, both background and bursts for Nambu-Goto strings \cite{Hogan:2006we}.
Gravitaional signals that can be detected by LISA $P-n$-$H-n$ is signal for cosmic strings with tension $G \mu \sim 10^{-n}$. Various other signals are also shown: MBHB is for massive black hole binary system etc.. The sensitivity of LISA is shown in red. So is that for milli-second pulsar timing (MSP) at low frequency. We see that LISA can reach $G\mu \sim 10^{-15}$.
}
\label{5.4}
\end{center}
\end{figure}
$\bullet$ As a cosmic string moves with velocity {\bf v} across the sky,
a shift in the CMB temperature may be observed,
$\Delta T/T \simeq 8 \pi G\mu v \gamma$ \cite{deltat}. 
A careful analysis of the CMBR 
data may probe $G\mu \simeq 10^{-10}$. It is important to see what 
bound on $G \mu$ the data can eventually reach. 
Detection may be possible for as small as $G\mu \simeq 10^{-13}$. \\
$\bullet$The cosmic string network also generates gravitational waves that may 
be observable. This has been studied extensively in the literature.
The stochastic gravitational wave spectrum has an almost flat region 
that extends 
from $f \sim 10^{-8}$ Hz to $f \sim 10^{10}$ Hz. 
Within this frequency range, both ADVANCED LIGO/VIRGO (sensitive at around 
$f \sim  10^2$ Hz) and LISA (sensitive at around $f \sim  10^{-3}$ Hz) may have a chance.
Following Ref.\cite{caldwell}, we obtain
$ \Omega_{gw} h^2 \simeq  0.04 G\mu$ coming from long strings.
Since LIGO II/VIRGO can reach
$\Omega_{gw} h^2 \simeq 10^{-10}$ at $f \simeq 100$ Hz,
it can reach $G\mu \ge 2 \times 10^{-9}$. Such stochastic gravitational 
wave also influences the very precise pulsar timing measurements.
Although present pulsar timing measurement is 
compatible with $G \mu < 10^{-6}$, a modest improvement on the 
accuracy may detect a network of cuspy cosmic string loops down to 
$G \mu \simeq 10^{-11}$. 

 Cusps and kinks are quite common in oscillating cosmic strings. 
Strongly focused beams of relatively high-frequency gravitational 
waves are emitted by these cusps and kinks.
The sharp bursts of gravitational waves have very
distinctive waveform: $t^{1/3}$ (cusps) and $t^{2/3}$ (kinks) \cite{Damour:2001bk}.
ADVANCED LIGO/VIRGO may detect them for values down to 
$G \mu \ge 10^{-13}$ and LISA to $10^{-15}$ \cite{Damour:2001bk,Siemens:2006vk,Hogan:2006we}, 
so this may be the most sensitive test of cosmic strings. At the moment, theoretical uncertainties
(such as string tension, tension spectrum, interactions and cosmic string loops) must be better understood. Figure \ref{5.4} takes into account the recent analysis where the string
loops are important. \\
$\bullet$ Cusps also introduces temperature shifts in the CMBR that should be searched.
They may appear as a sharp down and then up temperature shift that is quite distinctive
\cite{deLaix:1996vc,chernoff}.

\section{Remarks}

Brane inflation is a natural realization of inflation in the brane 
world scenario in string theory. If the string scale is close to the GUT 
scale, as expected, cosmology offers a powerful approach to study and test string theory.
We see that brane inflation offers a variety of possible distinct stringy signatures 
to be detected. Existing data is perfectly compatible with brane inflation. It is 
exciting that near future experiments/observations will likely 
provide non-trivial tests of the scenario.

Many interesting problems remain. Here is a partial list. On the theoretical side:  \\
$\bullet$ Search for other inflationary scenarios in string theory. \\
$\bullet$ Search for other distinct stringy signatures that can be detected. \\
$\bullet$ We have seen that the structure of the bulk as well as the properties of the 
warped deformed throat impacts on the CMBR predictions, e.g., the power spectral index.
Flux compactifications must be studied in much greater details than currently known. \\
$\bullet$ The gauge/gravity duality has played an important role in studying the properties 
of throats and the cosmic string tension spectrum. One may actually apply cosmology to study
strongly coupled gauge theory via gauge/gravity duality. \\
$\bullet$ Non-Gaussianity in CMBR and its more detailed properties. \\
$\bullet$ Understand better the properties of cosmic strings, such as the tension spectrum and their interactions, their production and stability, and the cosmological evolution of the string 
network that may include baryons and/or light domain walls bounded by the cosmic strings. \\
$\bullet$ Gott finds that closed timelike curves appear when two cosmic strings move 
ultra-relativistically towards each other \cite{Gott:1990zr}. He proposed to use this as a time 
machine. However, a photon will be instantly blue-shifted to infinite energy too. It is argued that energetics would prevent the appearance of such closed timelike curves in our universe under any realistic situations \cite{Shlaer:2005gk}. That is, photons in CMBR would simply sap enough 
energies from the cosmic strings to prevent the appearance of closed timelike curves.
This important and fascinating issue certainly deserves further analysis. \\
On the observational side :\\
$\bullet$ Searching for cosmic string signatures, large tensor mode and/or non-Gaussianity 
that differs from that predicted in slow-roll inflation in CMBR will be important. \\
$\bullet$  Astronomical searches for lensing, micro-lensing, temperature shifts due to 
moving strings and string cusps can be both challenging and exciting. Some of these 
searches need not be dedicated searches, that is, they can be part of other programs. \\
$\bullet$ Gravitational wave detection of the stochastic background gravitational radiation 
due to cosmic strings as well as bursts coming from string cusps will be valuable. 

One should consider the discovery of cosmic strings as another verification of the inflationary paradigm. 
This will shed light on the specific brane inflationary scenario that took place, providing a 
valuable probe to the brane world picture before inflation. 
That is, information on the early universe before inflation may not be totally lost. 
To my knowledge, this is the best observational window into supertstring theory.
Irrespective of the final outcome, whether brane inflation or some other stringy scenario is 
eventually proven correct or not, we see that string theory is confronting data and making 
a number of qualitatively as well as quantitatively distinctive predictions that can be tested 
in the near future. This is exciting.

\vspace{5mm}

{\bf Acknowledgment}


\vspace{3mm}

I thank Rachel Bean, Xingang Chen, David Chernoff, Gia Dvali, Hassan Firouzjahi, 
Girma Hailu, Nick Jones, Louis Leblond, Levon Pogosian, Sash Sarangi, Sarah Shandera, Gary Shiu, Ben Shlaer, Horace Stoica, Ira Wasserman, Mark Wyman and Jiajun Xu 
for collaborations and valuable discussions.
Discussions with Cliff Burgess, Jim Cline,
Shamit Kachru, Renata Kallosh, Igor Klebanov,
Andre Linde, Liam McAllister, Juan Maldacena, Irit Maor, Ken Olum, Joe Polchinski, 
Fernando Quevedo, Eva Silverstein, 
Bret Underwood and Alex Vilenkin are gratefully acknowleged.
This work is supported by the National Science Foundation under grant PHY-0355005.

\end{document}